\documentclass{article}
\usepackage{spconf,amsmath,graphicx}

\usepackage{url}
\usepackage{color}

\usepackage{pifont}

\usepackage{array}
\newcolumntype{P}[1]{>{\centering\arraybackslash}p{#1}}
\newcolumntype{M}[1]{>{\centering\arraybackslash}m{#1}}

\usepackage{multirow}
\usepackage{amsmath}
\usepackage{amssymb}
\usepackage[table]{xcolor}
\usepackage{booktabs}

\usepackage{graphicx}
\usepackage{hyperref}

\usepackage[ruled,linesnumbered]{algorithm2e}

\title{Low-Complexity Acoustic Echo Cancellation with
Neural Kalman Filtering}
\name{Dong Yang$^*$, Fei Jiang$^*$, Wei Wu, Xuefei Fang, Muyong Cao\thanks{$^*$The first two authors contributed equally to this work.}}
\address{Tencent GVoice}
\begin{document}
%
\maketitle
\begin{abstract}
The Kalman filter has been adopted in acoustic echo cancellation due to its robustness to double-talk, fast convergence, and good steady-state performance. The performance of Kalman filter is closely related to the estimation accuracy of the state noise covariance and the observation noise covariance. The estimation error may lead to unacceptable results, especially when the echo path suffers abrupt changes, the tracking performance of the Kalman filter could be degraded significantly. In this paper, we propose the neural Kalman filtering (NKF), which uses neural networks to implicitly model the covariance of the state noise and observation noise and to output the Kalman gain in real-time. Experimental results on both synthetic test sets and real-recorded test sets show that, the proposed NKF has superior convergence and re-convergence performance while ensuring low near-end speech degradation comparing with the state-of-the-art model-based methods. Moreover, the model size of the proposed NKF is merely 5.3 K and the RTF is as low as 0.09, which indicates that it can be deployed in low-resource platforms. 
\end{abstract}
\section{Introduction}
\label{sec:intro}
Acoustic echo cancellation (AEC) plays an important role in the speech processing field. It aims to remove the echoes introduced by the far-end signals in the speech communication scenario, according to the far-end reference signals and microphone signals. Model-based (MB) methods (adaptive filters) such as least-mean square (LMS), normalized least-mean square (NLMS), and recursive least-squares (RLS) are widely used in this field. 

The state-space models such as Kalman filtering have also been successfully applied in echo cancellation \cite{FDKF-2006, KF-2010, FDKF-2012, FDKF-epc}. Comparing with other adaptive filters, Kalman filter is more robust to double-talk and has better convergence performance. Because it assumes that the echo path is time-varying, which is modeled by the state equation. The convergence and tracking ability of Kalman filtering are controlled by the covariance of the state noise and observation noise, which can be dynamically estimated according to some assumptions. However, the noise covariance estimation error may lead to unacceptable results. For instance, Yang \textit{et al.} \cite{FDKF-epc} pointed out that the overestimation of the observation noise covariance can degrade the reconvergence speed when the echo path changes abruptly. 

In order to solve this problem, we propose to combine the MB method and data-driven (DD) method. More specifically, we propose to use a trainable RNN to implicitly model the covariance of the state noise and observation noise and output the adaptive Kalman gains in real-time. This can result in better convergence and tracking ability than the Kalman filter \cite{KN1, KN2}. 
Unlike other neural network aided adaptive filtering methods, e.g., deep neural network-controlled frequency-domain adaptive filter (DNN-FDAF) \cite{DNN-FDAF} and deep variable step-size NLMS (DVSS-NLMS) \cite{DVSS-NMLS} that respectively filtering in the frequency domain and in the time domain,
we formulate the neural Kalman filtering operations in the time-frequency domain, resulting in a much smaller model size (with merely 5.3 K comparing to 2.4 M and 1 M parameters). 
Experimental results on both synthetic test sets and real-recorded test sets show that, the proposed NKF achieves much better AEC performance than the traditional adaptive filters and the state-of-the-art hybrid MB/DD method Meta-AF \cite{Meta-AF}. Moreover, comparing with the deep-learning based AEC methods that are fully data-driven \cite{Zhang18_interspeech, CAD-AEC, DTLN-AEC, F-T-LSTM}, the proposed method has much smaller model size, and introduces less near-end signal distortion according to the experimental results. 
The inference code and the pre-trained model are released online~\footnote{Code available at \url{ https://github.com/fjiang9/NKF-AEC}}.

\section{Proposed Method}
\subsection{AEC Signal Models}
\label{sec:obs_model}
In acoustic echo cancellation, the microphone signal $y[n]$ is assumed to be the summation of the near-end signal $s[n]$ and the echo of the far-end signal $x[n]$, which depends on the room impulse response $h[n]$:
\begin{align}
    y[n] = s[n] + \sum_{\tau=0}^{T-1}h[\tau]x[n-\tau],
\label{eq:mic_t}
\end{align}
where $T$ is the length of the room impulse response (RIR) $h[n]$. The second term in Eq.~\eqref{eq:mic_t} is the linear acoustic echo. Note that there are also some non-linear components in realistic scenarios, but we only consider the linear echo in this paper. The objective of AEC is to estimate $h[n]$ so that the echo-cancelled signal can be obtained by
\begin{align}
    \hat s[n] = y[n] - \sum_{\tau=0}^{T-1}\hat h[\tau]x[n-\tau].
\label{eq:aec_t}
\end{align}

Similar to Ref.~\cite{SAEC-MLE}, we approximately reformulate Eq.~\eqref{eq:mic_t} into the time-frequency (TF) domain by applying short-time Fourier transform (STFT) as
\begin{align}
    Y[m, k] \approx S[m, k] + \sum_{l=0}^{L-1}H^*[l, k]X[m-l, k],
\label{eq:mic_tf}
\end{align}
where $S[m, k]$, $Y[m, k]$, and $X[m, k]$ are respectively the near-end signal, microphone signal, and far-end signal at the frame $m$ and frequency $k$, $L$ is the number of taps, and $^*$ refers to the complex conjugate operation. $H[l, k]$ is the convolutive transfer function (CTF) of the echo path. To simplify the expression, we define 
\begin{align}
    \mathbf x_{m,k} &= \left[X[m, k], \cdots, X[m-L+1, k]\right]^\text T, \\
    \mathbf{h}_k &= [H[0, k], \cdots, H[L-1, k]]^\text{T}.
\label{eq:x_h}
\end{align}
Then the echo-cancelled signal can be expressed as
\begin{align}
    \hat S[m, k] = Y[m, k] - \hat{\mathbf h}_k^{\text H}\mathbf x_{m,k},
\label{eq:s_hat}
\end{align}
where $^{\text H}$ stands for the Hermitian transpose.

\subsection{Kalman Filtering in the TF domain}
Most previous Kalman filtering adopted in AEC are processed either in the frequency domain~\cite{FDKF-2006, PBFDKF-2014, FDKF-epc} or in the time domain~\cite{TDKF-2013}. In this paper, we introduce Kalman filtering to the TF domain for echo cancellation and denote it as TF-domain Kalman filter (TFDKF). 

The echo path $\mathbf{h}_k \in \mathbb{C}^{L}$ can be assumed as a zero-mean random vector, which follows a simplified first-order Markov model, i.e.,
\begin{align}
    \mathbf{h}_{m,k} = A\mathbf{h}_{m-1,k} + \mathbf{w}_{m,k},
\label{eq:hk_state}
\end{align}
where $A$ is a transition parameter \cite{FDKF-2006} that is close to 1, $\mathbf{w}_{m,k} \in \mathbb{C}^{L}$ is a noise vector, and $\mathbf{w}_{m,k} \sim \mathcal N(0, \mathbf Q_k)$. Eq.~\eqref{eq:hk_state} is called the state equation. The covariance matrix $\mathbf Q_k$ can determine the tracking ability of the Kalman filter. We denote the correlation matrix of the prior misalignment $\mathbf{h}_{m,k}-\hat{\mathbf{h}}_{m-1,k}$ as $\mathbf{P}_{m,k}$, i.e., $\mathbf{P}_{m,k}=E[(\mathbf{h}_{m,k}-\hat{\mathbf{h}}_{m-1,k})(\mathbf{h}_{m,k}-\hat{\mathbf{h}}_{m-1,k})^{\text H}]$, then
\begin{align}
    \mathbf{P}_{m,k} = A^2\mathbf{P}^+_{m-1,k} + \mathbf Q_k,
\label{eq:P_state}
\end{align}
where $\mathbf{P}^+_{m,k}$ represents the correlation matrix of posterior misalignment $\mathbf{h}_{m,k}-\hat{\mathbf{h}}_{m,k}$. The observation function can be expressed as 
\begin{align}
    Y[m,k] = \mathbf h_{m,k}^{\text H}\mathbf x_{m,k} + S[m,k].
\label{eq:P_obs}
\end{align}
The Kalman filter solution can be expressed as 
\begin{align}
    \hat{\mathbf{h}}_{m,k} &= A\hat{\mathbf h}^+_{m-1,k}, \label{eq:kf_process1}\\ 
    \mathbf{P}_{m,k} &= A^2\mathbf{P}_{m-1,k}^+ + \mathbf Q_k, \\
    e_{m,k} &= Y[m,k] - \hat{\mathbf h}_{m,k}^\text{H}\mathbf x_{m,k}, \label{eq:prior_e}\\
    \mathbf{k}_{m,k} &= \mathbf P_{m, k}\mathbf x_{m,k}\left(\mathbf x_{m,k}^{\text H}\mathbf P_{m, k}\mathbf x_{m,k}+|S[m,k]|^2\right)^{-1}, \\
    \hat{\mathbf{h}}^+_{m,k} &= \hat{\mathbf h}_{m,k} + \mathbf{k}_{m,k}e_{m,k}, \\ 
    \mathbf{P}_{m,k}^+ &= \left[\mathbf I_L-\mathbf{k}_{m,k}\mathbf x_{m,k}^{\text H}\right]\mathbf{P}_{m,k},
\label{eq:kf_process}
\end{align}
where $\mathbf k_{m,k}\in \mathbb{C}^{L}$ is the Kalman gain. In practice, 
$\mathbf Q_k$ can be estimated using different schemes \cite{KF-2010, TDKF-2013, wu2016robust}, we estimate it by
\begin{align}
    \hat{\mathbf Q}_k = (1-A^2)E[\hat{\mathbf h}_{m-1,k}\hat{\mathbf h}_{m-1,k}^{\text H}].
\label{eq:q_k}
\end{align}
The observation noise $S[m,k]$ can be estimated by the prior error signal $e_{m,k}$ according to Eq.~\eqref{eq:prior_e}.

\subsection{Neural Kalman Filtering for AEC}
Because the estimation of $|S[m,k]|^2$ and $\mathbf Q_k$ are generally not accurate, the Kalman gain is therefore not optimal. For instance, when an abrupt echo path change occurs, the power of the prior error signal $e_{m,k}$ could be quite large, thus leading to overestimation of the observation noise covariance. As a result, the reconvergence speed will be very slow according to the analysis in \cite{FDKF-epc}.

To address the problem of noise statistics estimation in Kalman filter, we propose to use a RNN model to implicitly model the noise statistics and directly estimate the Kalman gain in real-time. From Eqs.~\eqref{eq:kf_process1}-\eqref{eq:kf_process} we can see that, the Kalman gain $\mathbf k_{m,k}$ is mainly related to the far-end signal $\mathbf x_{m,k}$, the noise covariance $\mathbf Q_k$, and the near-end signal power $|S[m,k]|^2$. $\mathbf Q_k$ and $|S[m,k]|^2$ can be inferred from the innovation of $\hat{\mathbf h}$ at the last frame and the prior echo-cancelled signal at the current frame, respectively. Therefore, we concatenate $\mathbf x_{m,k}$, $\Delta\hat{\mathbf h}_{m-1,k}=\mathbf{k}_{m-1,k}e_{m-1,k}$, and $e_{m,k}$ as the input features $\mathbf z_{m,k} \in \mathbb{C}^{D}$ to output the Kalman gain $k_{m,k}$. Obviously, the input feature dimension $D$ is related to the taps $L$, i.e., $D=2L+1$.

The model consists of one complex fully-connected (FC) layer, two complex GRU layers, and another two complex FC layers. The PReLU activation function \cite{prelu} is used after the first two FC layers. The numbers of units in the three FC layers are $2D$, $2D$, and $L$ respectively. The number of units in the two GRU layers are $L^2+2$. Interestingly, we empirically find that increasing the number of hidden units in the GRU layers could only marginally improve the AEC results. This is reasonable because the dimension $L^2+2$ is sufficient for modeling both $\mathbf Q_k$ and $|S[m,k]|^2$. It is noted that the RNN model independently runs at each frequency bin, which means that the frequency dimension of the input features can be viewed as the batch dimension. This design can greatly decrease the number of parameters of the model as the feature dimension is only at the $L$ level and the networks can share among different frequencies.

The forward pass of neural Kalman filtering at the $k$-th frequency bin is shown in Algorithm \ref{alg:nkf}. The initialization of echo path $\mathbf h_{0,k}$, RNN hidden state $\mathbf g_{0,k}$, and $\Delta \mathbf h_{0,k}$ can be either all-zero vectors or white Gaussian noises. Using an all-zero vector to initialize the echo path $\mathbf h_{0,k}$ is quite common in most previous research. The reason why we use white Gaussian noise as the initialization is to simulate the system state when an abrupt echo path change occurs. We empirically set the probability of zero initialization and noise initialization as both 50\% during training, while keeping using zero initialization in the inference stage. After one forward pass, the estimated echo $\hat D[m,k]$ and echo-cancelled signal $\hat S[m,k]$ can be obtained. The loss function can then be calculated by measuring the distance between the estimated echo $\hat D[m,k]$ and ground-truth echo $D[m,k]$, e.g., 
\begin{align}
    \mathcal L(D, \hat D|\Theta) = \sum_{k=1}^{F}\sum_{m}|D[m,k]-\hat D[m,k]|^2.
\label{eq:loss}
\end{align}
Finally, The RNN model $\mathcal F_{\Theta}(\cdot,\cdot)$ is optimized using the gradient descent algorithm. In the inference time, the time-domain echo-cancelled signal $\hat s[n]$ can be reconstructed according to $\hat S[m,k]$ using inverse STFT.

\begin{algorithm}[t]
\caption{NKF forward pass at frequency $k$}\label{alg:nkf}
\KwIn{mic signal $Y$, far-end signal $\mathbf x$, model $\mathcal F_{\Theta}$}
\KwOut{echo-cancelled signal $\hat S$}
$\text{Initialize} ~ \hat{\mathbf h}_{0, k}, \mathbf g_{0, k}, \text{and} ~ \Delta\hat{\mathbf{h}}_{0,k}$\;
\For{$m \leftarrow 1$ \KwTo $T$}
{
$e_{m,k} = Y[m,k] - \hat{\mathbf h}_{m-1,k}^\text{H}\mathbf x_{m,k}$\;
$\mathbf z_{m, k} = [\mathbf x_{m,k}^{\text T}, \Delta\hat{\mathbf{h}}_{m-1,k}^{\text T}, e_{m,k}]^{\text T}$\;
$\mathbf k_{m,k}, \mathbf g_{m, k} = \mathcal F_{\Theta}(\mathbf z_{m,k}, \mathbf g_{m-1, k})$\;
$\Delta\hat{\mathbf{h}}_{m,k} = \mathbf{k}_{m,k}e_{m,k}$\;
$\hat{\mathbf h}_{m, k} = \hat{\mathbf h}_{m-1,k} + \Delta\hat{\mathbf{h}}_{m,k}$\;
$\hat S[m,k] = Y[m, k] - \hat{\mathbf h}_{m,k}^{\text H}\mathbf x_{m,k}$\;
}
\end{algorithm}

\section{Experiments}
\label{sec:exp}

\subsection{Experimental Setup}
We use the far-end speech in the synthetic fold of the AEC challenge \cite{icassp2021_aec} as the training corpus, which contains 10000 speech clips and each clip is 10 seconds long. During training, both the far-end signal and near-end signal are randomly sampled from this speech corpus. We use randomly generated white Gaussian noise as the RIR to generate the far-end echo signal. The signal-to-echo ratio (SER) is uniformly sampled from -5 dB to 5 dB. The duration of the far-end signal is randomly clipped to 1 s, while the near-end signal is randomly segmented to 0.5$\sim$1 s. 
The sampling rate of the audio is 16 kHz in this paper. As for the STFT parameters, the FFT size window length, and hop size are 1024, 1024, and 256, respectively. In this paper, the number of taps of the NKF is set to be 4 (i.e., $L=4$), which results in only 5.3 K real-valued trainable parameters. 

Two test sets are used to evaluate the AEC performance of the proposed method. One is a synthetic test set, which uses the speech data from the Librispeech~\cite{LBS} \textit{dev-clean} and \textit{test-clean} folds as the and the far-end and near-end signals respectively, and uses simulated RIRs in \cite{RIRs-Noises} to generate echo signals. Note that these data are not seen during training. The echo paths are cut off after 64 ms. The SER is uniformly sampled from -10 dB to 10 dB. 
The duration of each samples is 8 seconds. Four subsets, namely far-end single-talk (FST), far-end single-talk with echo-path change (FST-EPC), double-talk (DT), and double-talk with echo-path change (DT-EPC) are generated. Each subset contains 500 samples. The echo path change is implemented by abruptly switching the RIR from one simulated room to another at a certain moment (randomly from 3.5 s to 4.5 s).
The other is a real recorded test set, i.e., the clean \textit{blind test set} from the ICASSP 2021 AEC Challenge \cite{icassp2021_aec}. 
We use the double-talk partition for testing, which contains 200 samples and half of them suffers echo path change. 
Note that some samples in this test set have large time delay, we therefore performs time-delay compensation using the GCC-PHAT algorithm \cite{gcc-phat} like in Ref. \cite{wRLS-DeepFSMN} before linear filtering.

\subsection{Results}

\begin{table*}[t]
\centering
\caption{Comparison of AEC methods with ERLE (dB) and PESQ across four subsets of the synthetic test set, and with AECMOS score for the real-recorded test set. The ERLE reported in this table is the mean of the segmental ERLEs. Model size (MS) refers to the number of real-valued parameters. Grey cells indicate that the method is not data-driven. 
}
\renewcommand{\arraystretch}{0.6}
\begin{tabular}{M{55pt} M{15pt} M{30pt} | M{30pt} M{45pt} | M{30pt} M{30pt} | M{30pt} M{30pt} | M{30pt} M{30pt}}
\toprule
&&& \multicolumn{6}{c}{\textbf{Synthetic}} & \multicolumn{2}{|c}{\textbf{Real-recorded}} \\\cmidrule(lr){4-9}\cmidrule(lr){10-11}
\textbf{Method} & \textbf{Lin.} & \textbf{MS} & \textbf{FST} & \textbf{FST-EPC} & \multicolumn{2}{c}{\textbf{DT}} & \multicolumn{2}{|c|}{\textbf{DT-EPC}} & \multicolumn{2}{c}{\textbf{DT (AECMOS)}}\\\cmidrule(lr){4-4}\cmidrule(lr){5-5}\cmidrule(lr){6-7}\cmidrule(lr){8-9}\cmidrule(lr){10-11} 
&&& ERLE & ERLE & ERLE & PESQ & ERLE & PESQ & Echo & Other \\
\midrule
PNLMS \cite{PNLMS} & $\checkmark$ & {\cellcolor{black!5} --} & 10.86 & 8.75 & 7.06 & 1.50 & 5.52 & 1.43 & 2.64 & 3.94 \\
PFDKF~\cite{PBFDKF-2014} & $\checkmark$ & {\cellcolor{black!5} --} & 25.87 & 17.56 & 7.46 & 1.46 & 5.88 & 1.41 & 2.75 & 3.78 \\
TFDKF & $\checkmark$ & {\cellcolor{black!5} --} & 24.50 & 18.62 & 15.11 & 2.29 & 10.99 & 1.77 & 2.89 & 3.96 \\
\midrule
DTLN~\cite{DTLN-AEC} & -- & 10.4 M & -- & -- & -- & -- & -- & -- & \textbf{4.31} & 3.73  \\
Meta-AF~\cite{Meta-AF} & $\checkmark$ & 28 K & 19.45 & 14.60 & 12.59 & 1.78 & 9.09 & 1.56 & 2.64 & 3.85 \\
NKF & $\checkmark$ & 5.3 K & \textbf{28.41} & \textbf{24.75} & \textbf{15.99} & \textbf{2.77} & \textbf{13.75} & \textbf{2.37} & 4.02 & \textbf{4.02}  \\
\bottomrule
\end{tabular}
\label{tab:syn_test_results}
\end{table*}

\begin{figure}[t]
\centering
\includegraphics[width=3.4in]{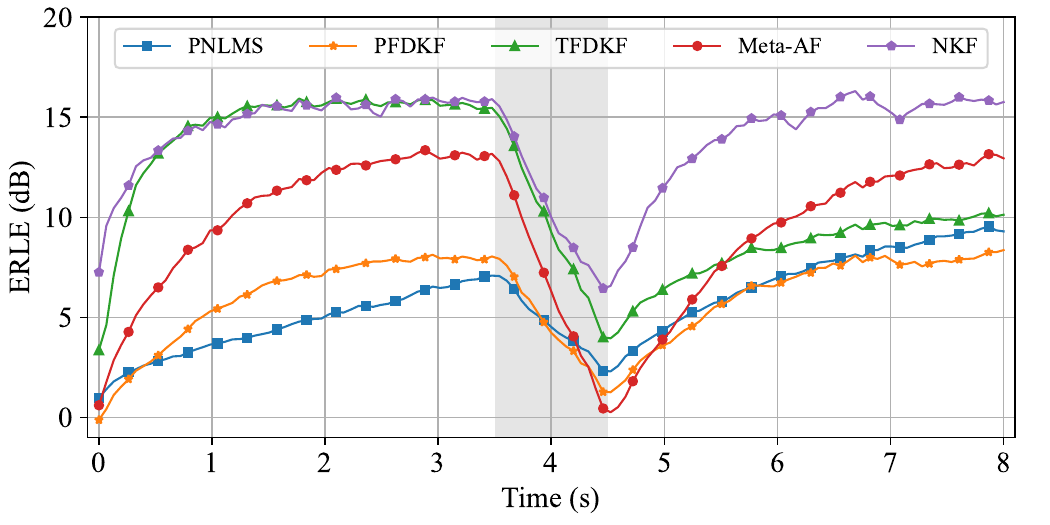}
\caption{ERLE curves on the synthetic DT-EPC subset. Abrupt echo path change occurs at the shaded region.}
\label{fig:erle}
\end{figure}

Three fully model-based methods, i.e., proportionate normalized least-mean-squares (PNLMS)~\cite{PNLMS}, partitioned-block-based frequency-domain Kalman filter (PFDKF)
~\cite{PBFDKF-2014}, and TFDKF, as well as one hybrid MB/DD method Meta-AF \cite{Meta-AF} are used as the baseline methods to compare with the proposed method. We directly use the pre-trained Meta-AF model released online by the authors. The hyper-parameters of the model-based baseline methods are tuned to be optimal upon the synthetic test set. For a fair comparison, the filter lengths of these linear AEC systems are set to be identical (i.e., 64 ms) in our experiments.

We use the segmental echo return loss enhancement (ERLE) \cite{ERLE} and wide-band perceptual evaluation of speech quality (PESQ) \cite{PESQ} to evaluate the AEC performance on the synthetic test set. 
The results are shown in Table \ref{tab:syn_test_results}. We can see that the ERLE improvements from the best baseline results on the FST, FST-EPC, DT, and DT-EPC subsets are 2.54 dB, 6.13 dB, 0.88 dB, and 2.76 dB, respectively. The PESQ improvements from the best baseline results on the DT and DT-EPC subsets reach 0.48 and 0.60 respectively, which means that the NKF can better suppress the echo without introducing extra near-end speech distortion. The improvements on the FST-EPC and DT-EPC subsets is more significant indicates that the proposed NKF can estimate the Kalman gain more accurately than its model-based counterpart TFDKF especially when the system suffers abrupt echo path changes.
Similar observations can also be obtained from Fig. \ref{fig:erle}, which shows the ERLE curves (averaged on 500 test samples) of different linear AEC methods on the DT-EPC subset. We can clearly see that, the proposed NKF has much faster convergence and re-convergence speed comparing with other methods. In particular, it is interesting to notice that the re-convergence speed of the TFDKF is much slower than its convergence speed, whereas the re-convergence speed of the NKF is still fast enough. 
Such excellent tracking ability of the proposed NKF may due to the dynamic estimations of the noise statistics by the hidden states of the RNN layer. In addition, the model size of the proposed NKF is only 5.3 K, which is the smallest neural AEC model to the best of our knowledge. Its real-time factor (RTF), which is tested on a single CPU core, is as low as 0.09. 

The AECMOS results on the ICASSP 2021 AEC challenge blind test set double-talk partitions are shown in the real-recorded section of Table \ref{tab:syn_test_results}. In addition to the linear AEC baselines, we also report the results of a fully data-driven model DTLN 
\cite{DTLN-AEC} which ranked in the 3-rd place in the AEC challenge for comparison. We directly use its official pre-trained model for testing.
Since there is no ground-truth signal provided in the AEC challenge blind test set, a non-intrusive perceptual objective metric named AECMOS~\cite{AECMOS} is used to evaluate the AEC performance on the real recorded test set. 
We can see that, the proposed NKF works surprisingly well on this real-recorded test set. It surpasses all the other linear AEC baselines by a large margin, especially in terms of echo suppression (with at least 1.13 AECMOS-echo improvement). This indicates that conventional linear filtering methods may need to carefully tune the algorithm parameters in different scenarios. By contrast, The proposed NKF does not require tuning once training finished and has excellent generalization ability. Although the echo suppression performance of NKF is not as good as that of DTLN, the near-end speech quality of NKF is significantly higher. Considering that the NKF only suppresses part of the linear echo, the results may be further improved with longer filter length and a residual echo suppression network.

\section{Conclusions}
In this paper, we propose a new linear AEC framework based on Kalman filtering and neural networks. It significantly improves the performance of the Kalman filters while keeps a very small model size and fast inference speed, which is feasible to be adopted in the low-resource applications. For future work, we will further investigate better architecture for modeling the Kalman gain and extend the proposed method to multi-channel and non-linear echo cancellation. 

\vfill\pagebreak



\bibliographystyle{IEEEbib}
\small\bibliography{mybib}

\end{document}